\documentclass[%
 reprint,
 amsmath,amssymb,
aps,
prb,
superscriptaddress
]{revtex4-2}

\usepackage{color}
\usepackage{graphicx}% Include figure files
\usepackage{dcolumn}% Align table columns on decimal point
\usepackage{bm}% bold math
\usepackage{amsmath}
\usepackage{amssymb}
\usepackage{float}
\usepackage{url}

\begin{document}

\preprint{APS/123-QED}

\title{Chiral Diffraction from Aperiodic Monotile Lattice}
% Force line breaks with \\

\author{Yuto Moritake}
\altaffiliation{moritake@phys.titech.ac.jp}
\affiliation{Department of Physics, Institute of Science Tokyo, 2-12-1 Ookayama, Meguro-ku, Tokyo 152-8550, Japan}
%\affiliation{PRESTO, Japan Science and Technology Agency, 4-1-8 Honcho, Kawaguchi, Saitama 332-0012, Japan}

\author{Masato Takiguchi}
\affiliation{NTT Basic Research Laboratories, NTT Corporation, 3-1 Morinosato-Wakamiya, Atsugi-shi, Kanagawa 243-0198, Japan}
\affiliation{Nanophotonics Center, NTT Corporation, 3-1, Morinosato-Wakamiya, Atsugi-shi, Kanagawa 243-0198, Japan}

\author{Takuma Aihara}
\affiliation{NTT Device Technology Laboratories, NTT Corporation, 3-1 Morinosato-Wakamiya, Atsugi-shi, Kanagawa 243-0198, Japan}

\author{Masaya Notomi}
\affiliation{Department of Physics, Institute of Science Tokyo, 2-12-1 Ookayama, Meguro-ku, Tokyo 152-8550, Japan}
\affiliation{NTT Basic Research Laboratories, NTT Corporation, 3-1 Morinosato-Wakamiya, Atsugi-shi, Kanagawa 243-0198, Japan}
\affiliation{Nanophotonics Center, NTT Corporation, 3-1, Morinosato-Wakamiya, Atsugi-shi, Kanagawa 243-0198, Japan}

\date{\today}% It is always \today, today,

\begin{abstract}
Aperiodic systems such as quasiperiodic systems exhibit unique properties different from periodic structures.
In 2023, Smith \textit{et al.} discovered a new aperiodic structure: a single-shaped tile that can only tile space aperiodically, known as an aperiodic monotile.
Although the aperiodic monotile possesses intriguing mathematical properties, its experimental investigation remains unexplored.
In this study, we report an experimental investigation of diffraction patterns from a monotile lattice using a nanophotonic platform.
We observed clear Bragg peaks, which is evidence of long-range order and a chiral structure of the diffraction patterns.
Furthermore, we found exotic behavior in circular polarization dependence, which cannot be observed in conventional quasiperiodic structures.
These findings establish the monotile lattice as a novel class of aperiodic systems, expanding the study of nonperiodic structures beyond conventional quasicrystals.
\end{abstract}

\maketitle

%\section{\label{sec:level1}Introduction}
Periodic systems have been the central focus of research across various disciplines, including physics and chemistry.
However, aperiodic systems have attracted significant attention due to their unique properties and phenomena that cannot be observed in periodic systems.
In mathematics, studies on structures that can only tile the plane aperiodically have a long history.
Penrose discovered a set of two different tiles, known as Penrose tiles, that can only form aperiodic tilings \cite{Penrose1974}. 
The Penrose tilings are classified as quasiperiodic structures which don't have periodicity but exhibit long-range order \cite{Suck2002}.
Mathematical studies on such quasi-periodic systems eventually led to the discovery of quasicrystals by Shechtman \cite{PhysRevLett.53.1951} and he received Nobel prize in chemistry in 2011.
This discovery had a groundbreaking impact on crystallography, fundamentally altering the definition of crystals.
The concept of quasi-periodic structures and quasicrystals has been applied to various fields.
In nanophotonics, the concept of quasi-periodicity have been introduced in photonic crystals, photonic quasicrystals\cite{Vardeny2013,Steurer_2007,Maciá_2012}, leading interesting photonic functionalities such as the realization of complete photonic bandgaps \cite{10.1063/1.124848,zoorob2000,PhysRevB.63.081105,PhysRevLett.101.073902} and laser oscillation \cite{PhysRevLett.92.123906,10.1063/1.1762705,Mahler2010,Vitiello2014}.
More recently, studies on metasurfaces with quasi-periodic arrangements \cite{PhysRevLett.115.205501,xu2024}, and topological photonic systems using quasi-periodicity \cite{arjas2024,PhysRevLett.109.106402} have highlighted photonics as a versatile platform for both fundamental research and practical applications of quasi-periodic structures.

\begin{figure}
\includegraphics[width=\linewidth]{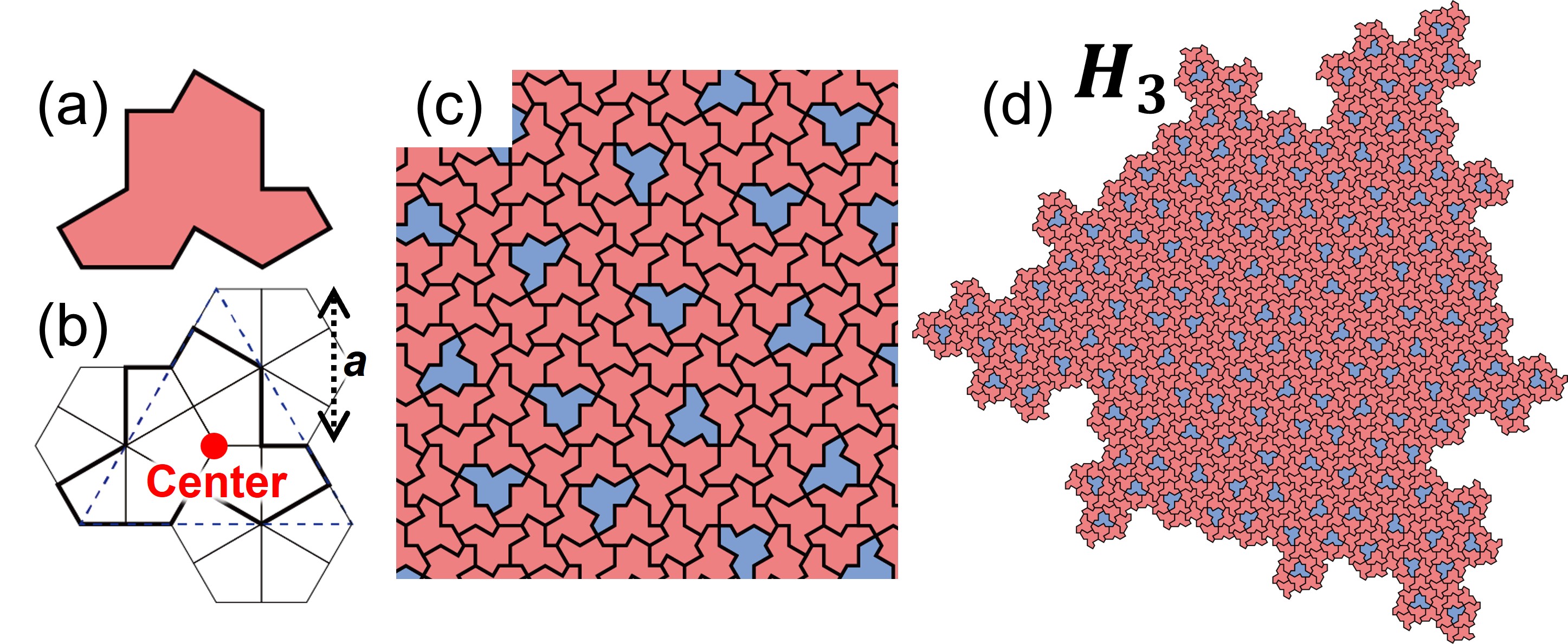}
\caption{\label{monotile}(a) The Smith hat tile. (b) Definition of the center of the Smith hat tile. (c) The hat tiling using the Smith hat tiles and its reversed tiles (illustrated in blue). (d) The $H_{3}$ metatile generated by the infulation rule.}
\end{figure}

%スミススタイルの発見
A natural mathematical question following the discovery of Penrose tiles is the existence of \textit{an aperiodic monotile}, a single shape that can tile the plane only non-periodically.
This problem, known as the "Einstein problem," remained unsolved for 50 years after the discovery of the Penrose tiles.
However, in 2023, Smith \textit{et al.} discovered this aperiodic monotile, attracting considerable attention \cite{smith2024}.
The first identified aperiodic monotile, resembling the shape of a hat, has been named the \textit{Smith hat tile} as shown in Fig \ref{monotile}(a).
Whether this new aperiodic tiling exhibits properties similar to quasiperiodic systems or possesses entirely new characteristics is an intriguing question.
No experimental study of this question has been conducted, although several studies for monotile-based systems \cite{PhysRevLett.132.086402,Kaplan:uv5021,Okabe_2024, JIN2024130781,PhysRevB.110.075435,PhysRevB.109.L220303,202407.1532,CLARKE2023101959} have been reported.

%It has been reported that this hat tile lattice exhibits quasicrystalline properties \cite{PhysRevB.108.224109}.
%Subsequently, the other monotile, a spectre tile, which does not require mirror symmetric tile in tiling has been discovered \cite{Smith2024b}.

In this study, we report experimental investigation of these properties by observing the diffraction pattern of a lattice generated from the tiling of the aperiodic monotiles.
In experiments, we fabricated the aperiodic pattern with the monotile lattice on SiN films, and observed light diffraction patterns. 
The observed diffraction patterns exhibited clear Bragg peaks indicating the long-range order of the monotile lattice, which is an experimental evidence of quasi-periodicity of the monotile lattice.
Furthermore, we discovered an intriguing dependence on incident circular polarization, reflecting the fact that the structure is chiral and lacks mirror symmetry.
Our results revealed new physical properties of the aperiodic monotile and open up a new direction of research of aperiodic systems, distinct from conventional quasi-periodic structures and quasicrystals.

%The patterns reflected the chirality of the tiling, and we confirmed the reversal of chirality when the structure was mirrored.
%The diffraction patterns perfectly matched with theory using the Fourier transform images of the monotile lattice.
%We found that the twisted angle of the hat tiling corresponds the tilting of the diffraction patterns, which induces the chiral diffraction patterns.
%We also measure circular polarization dependence and found interesting behavior.
%Our results demonstrates that photonic structures are a valuable tool for studying and utilizing physical phenomena in systems based on a aperiodic monotile.

Smith discovered the monotile from the honeycomb framework (Fig. \ref{monotile}(b)). 
As illustrated in Fig. \ref{monotile}(c), it is possible to generate an aperiodic tiling using the hat tile and its inverted structure (illustrated by a blue tile).
Hereafter, we call this tiling as "hat tiling."
Although the hat tiling employs two types of tiles related by space inversion, unlike the Penrose tiling, these tiles share the same shape.
Therefore, the hat tile is considered the first discovered aperiodic monotile.
Smith's team subsequently discovered a truly single-type aperiodic monotile \cite{Smith2024b}, referred to as the specter tile.
In this paper, we define a pseudo-period $a$ of the hat tiling by the period of the base honeycomb frame (Fig. \ref{monotile}(b)). 

The hat tiling can be generated by using several substitution rules \cite{smith2024}. 
In this study, we use the inflation rule using four basic metatiles, $H$, $P$, $T$, and $F$ \cite{smith2024}.
In this inflation rule, the $H_{1}$, $P_{1}$, $T_{1}$, and $F_{1}$ metatiles are generated by combining the basic metatiles and $n$-th generation of the metatiles, $H_{n}$, $P_{n}$, $T_{n}$, and $F_{n}$ can be generated by the $H_{n-1}$, $P_{n-1}$, $T_{n-1}$, and $F_{n-1}$ metatiles (See S1 in Supplementary Information).
Figure \ref{monotile}(d) shows the $H_{3}$ metatile.
A distinctive feature of the hat tiling is that the metatiles are arranged with a twist relative to the base honeycomb frame (see S1 in Supplementary Information).
This twisting introduces the golden mean $\phi = \left( 1+\sqrt{5} \right)/2$ into the inflation factor, which does not typically appear in the honeycomb frame.
Furthermore, this twist plays a crucial role in determining the chirality of the hat tiling as discussed later.
In this paper, "chirality" refers to the absence of mirror symmetry within a two-dimensional plane, which is distinct from the conventional chirality defined in three-dimensional space.

\begin{figure}
\includegraphics[width=1\linewidth]{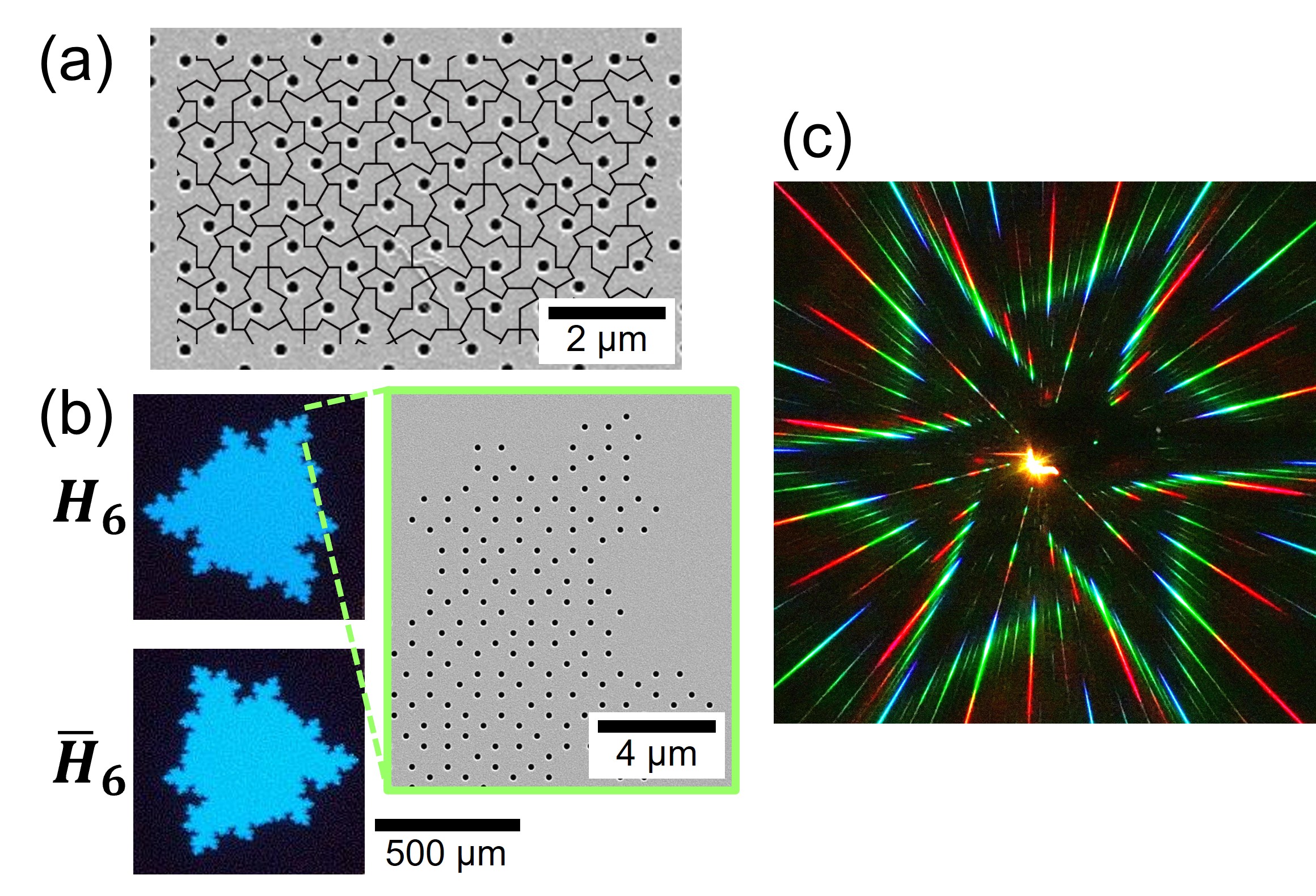}
\caption{\label{exp}(a) The monotile lattice generated by the center of the hat tile in the hat tiling. The SEM image of the monotile lattice is overlaid with the hat tiling. (b) Microscope images of the fabricated monotile lattice (left). A SEM image of the air holes on a SiN film (right). (c) A observed diffraction pattern from the monotile lattice patterned on the SiN film when a white light source was injected.}
\end{figure}

To generate the lattice from the hat tiling, we defined the lattice point at "the center" of each hat tile.
Although the hat tile has no specific symmetry, there is the center point (Fig. \ref{monotile}(b)), and this center is on the vertices of the base honeycomb frame.
The generated aperiodic lattice, the monotile lattice, is shown in Fig. \ref{exp}(a).
The hat tiling does not possess rotational symmetry, but the monotile lattice has $C_{3}$ rotational symmetry.
$C_{3}$ symmetry is allowed in conventional crystals, making the monotile lattice distinct from typical quasicrystals.
Furthermore, unlike quasicrystals such as the Penrose lattice, it is a chiral structure that lacks mirror planes, which gives rise to chiral diffraction patterns as shown later.

\begin{figure}
\includegraphics[width=\linewidth]{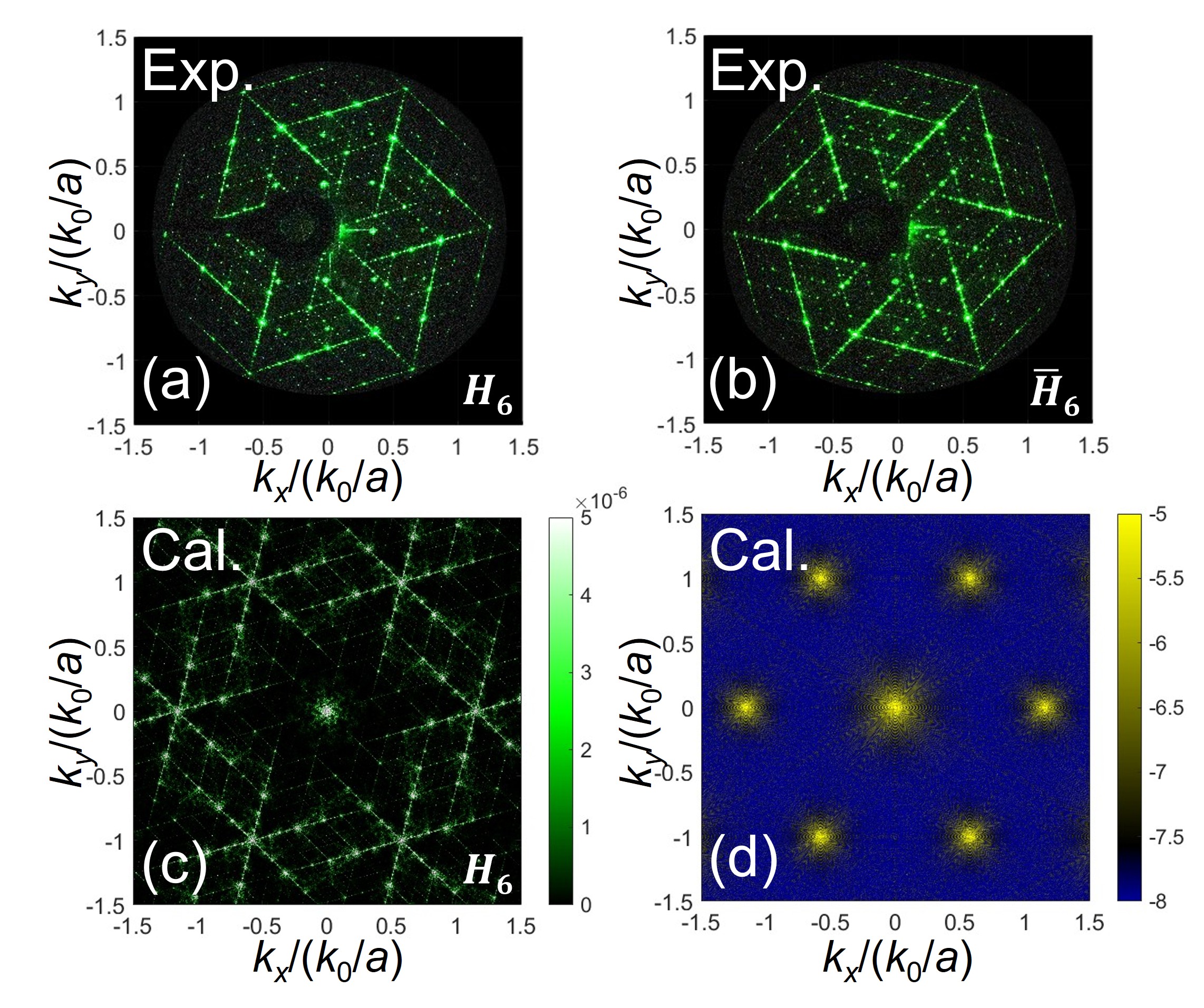}
\caption{\label{compare}Measured diffraction patterns of (a) $H_{6}$ and (b) $\bar{H_{6}}$ monotile lattices for green laser illumination. (c) Calculated diffraction using Fourier transform of the $H_{6}$ monotile lattice. (d) Logarithmic plot of calculated diffraction patterns of the honeycomb lattice. }
\end{figure}
%(e) Measeured diffration pattern of $H_{6}$ monotile lattice captured in high sensitivity mode. (f) Logarithmic plot of calculated diffraction patterns of $H_{6}$ monotile lattice.

For experiments, we fabricated monotile lattices using patterned SiN films.
A 350-nm-thick SiN film deposited on a Si substrate was patterned using electron beam lithography and etching to create circular holes with a radius of 100 nm and a depth of 350 nm at the $H_{6}$ monotile lattice points.
We prepared several devices with pseudo-period $a$ ranging from 600 to 750 nm.
To investigate chiral nature of the monotile lattice, inverted structures of $H_{6}$, $\bar{H_{6}}$, were also fabricated.
In the inverted hat tiling, the entire structure is reversed, resulting in a hat tiling predominantly composed of blue hat tiles.
Typical microscope and scanning microscope images of the fabricated monotile lattices are shown in Fig. \ref{exp}(b).
In the case of $H_{6}$, the number of the lattice points is 372100 and its area is approximately 500 $\mu m$ $\times$ 500 $\mu m$.

In diffraction experiments, a laser was normally incident to the sample surface, and the diffraction pattern in the reflected direction was projected onto a white screen and captured with a camera (See S2 in Supplementary Information).
The Fig. \ref{exp}(c) shows the diffraction pattern for the case of injection of a white light source (a supercontinuum laser).
A distinctive diffraction pattern with a pinwheel-like shape having clear chirality was observed.
This structure, being inherently two-dimensional, does not exhibit chirality originating from a three-dimensionality.
However, the twisting of the metatile tiling induces a global rotation of the entire structure, and this directionality manifests as chirality in the diffraction pattern.
Figure \ref{compare}(a) shows the diffraction pattern captured from normal direction of the screen when a laser with a wavelength of $\lambda =$ 532 nm.
Here, to clarify the diffraction patterns in wavenumber space, the real-space axes of the captured images were transformed into wavenumber axes through a coordinate transformation (See S2 in Supplementary Information).
The waveneumber axis is normalized  by the pseudo-period $a$ and vacuum wavenumber $k_{0} = 2\pi/\lambda$ for direct comparizon between experiments and calculations.
Numerous sharp diffraction peaks were clearly observed, indicating that the long-range order of the monotile lattice.
The diffraction pattern exhibits six-fold symmetry reflecting the $C_{3}$ symmetry of the monotile lattice.
Moreover, the diffraction patterns obtained from mirrored structures exhibited a reversal of chirality, as shown in Fig. \ref{compare}(b).

To analyze the observed diffraction patterns, theoretical calculations were performed.
By placing delta functions at the lattice points and performing a Fourier transform, diffraction patterns from the lattice can be calculated \cite{Kaplan:uv5021}.
Here, the Fourier amplitude of the diffraction patterns is written as
\begin{equation}
    F (k_{x}, k_{y}) = \sum_{i} \text{exp} \left[ 2 \pi i \left( k_{x} x_{i} + k_{y} y_{i} \right) \right]
\end{equation}
$k_{x}$ and $k_{y}$ are the wavevectors, and $x_{i}$ and $y_{i}$ are the coordinates of the lattice point.
Diffraction intensity is obtained by $I = |F|^2$.
We computed $I$ at each $k$-point for the monotile lattice of $H_{6}$ metatile as shown in Fig. \ref{compare}(c), which reproduced the observed diffraction pattern (Fig. \ref{compare}(a)) well.
Logarithmic plot and wide wavenumber view of Fig. \ref{compare}(c) are presented in S3 in Supplementary Information.
Not only the position of the diffraction peaks but also relative intensity is agree well each other.

%By using a high sensitivity mode of the camera, we can visualize the little peaks in the diffraction as shown in Fig. \ref{compare}(e) while strong peaks saturate.
%The detail structures of the observed peaks agree well with the calculation (Fig. \ref{compare}(f)).

The monotile lattice can be considered as the honeycomb lattice with lattice defects (See S1 and S3 in Supplementary Information).
Therefore, the diffraction pattern from the monotile lattice contains features characteristic of the diffraction pattern from the honeycomb lattice.
Figure \ref{compare}(d) is a calculated diffraction pattern for the honeycomb lattice in a logarithmic scale.
In this case, only sixfold-symmetric diffraction peaks appear.
The centers of the asterisk-like shape in the diffraction pattern of the monotile lattice (Fig. \ref{compare}(c)) coincides with the positions of the six-fold diffraction peaks of the honeycomb lattice (Fig. \ref{compare}(d)).

\begin{figure}
\includegraphics[width=\linewidth]{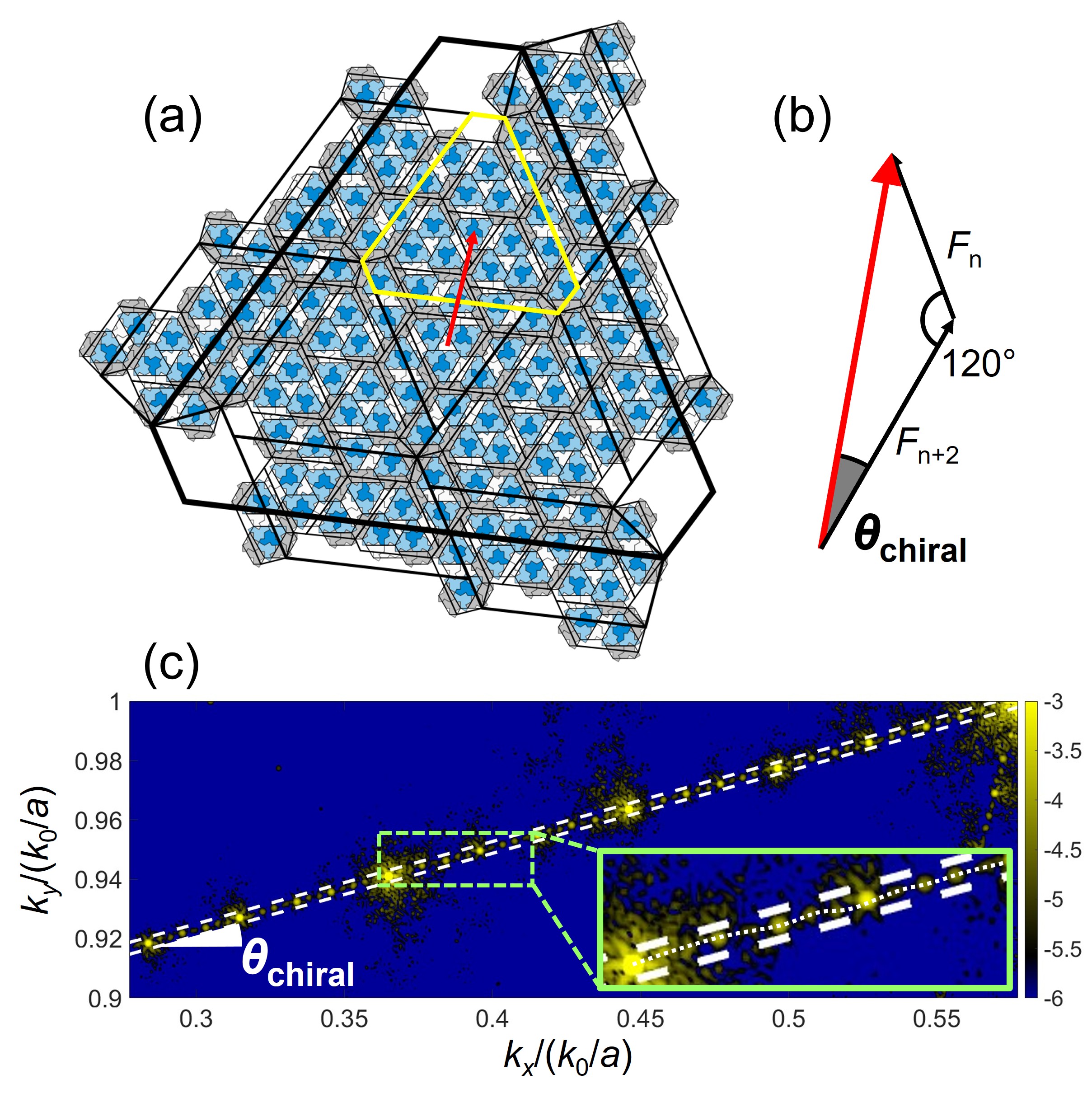}
\caption{\label{angle}(a) Twisting of the hat tiling as increase in the number of infulation. The red arrow is the vector from the origin to the center of the one of the $H$ metatile colored by yellow. The picture of the hat tiling is generated by a browser-based visualization tool provided by Smith's team (https://cs.uwaterloo.ca/\%7Ecsk/hat/). (b) The triangle to calculate $\theta_{\text{chiral}}$. (c) A magnified view of the asterisk-like shape in a logarithmic plot of Fig. \ref{compare}(c).}
\end{figure}

The origin of the chiral pinwheel structure in the diffraction patterns is tilting of the asterisk-shape structure.
We found that this angle $\theta_{\text{chiral}}$ can be analytically calculated from the twisting angle inherent in the monotile lattice.
As shown in Fig. \ref{angle}(a), the metatiles of the hat tiling gradually tilt away from the underlying honeycomb frame as the inflation rule is applied (See S1 in Supplementary Information).
This twisting is depicted by the tilting  of the red arrow in Fig. \ref{angle}(a).
The tilting angle $\theta_{\text{chiral}}$ can be calculated by the triangle as shown in Fig. \ref{angle}(b).
This triangle is composed of sides whose lengths correspond to the $n$-th and $n+2$-th terms of the Fibonacci sequence.
In the limit of $n \xrightarrow{} \infty$, their ratio can be expressed by the golden mean $\phi$
(See S1 in Supplementary Information).
The angle $\theta_{\text{chiral}}$ calculated using the triangle is given by
\begin{equation}
    \theta_{\text{chiral}} = \arccos{\frac{3\phi-1}{4}} \sim 15.52^{\circ}.
\end{equation}
Figures \ref{angle}(c) is a magnified view of the asterisk-shape region in Fig. \ref{compare}(f).
The white lines are drawn at equal intervals from the center of the asterisk with an angle of $\theta_{\text{chiral}}$.
It is evident that the diffraction peaks forming the asterisk align with the analytically derived angles $\theta_{\text{chiral}}$.
The geometric features of the monotile lattice in real space are directly reflected in the structure of wavevector space.
Closer inspection of the diffraction pattern as shown in the inset of Fig. \ref{angle}(f) (the light-green rectangle) reveals that the relatively weak peaks do not align strictly along straight lines but instead exhibit a meandering arrangement.
This feature is observed in the diffraction patterns for other generation of $H$ metatiles.
Moreover, we can find self-similar structures characterized by the gonlden mean $\phi$, which is observed in Fibonacci lattice and other quasi-periodic lattices \cite{photonics10091045} (See S3 in Supplementary Information).

\begin{figure*}
\includegraphics[width=\linewidth]{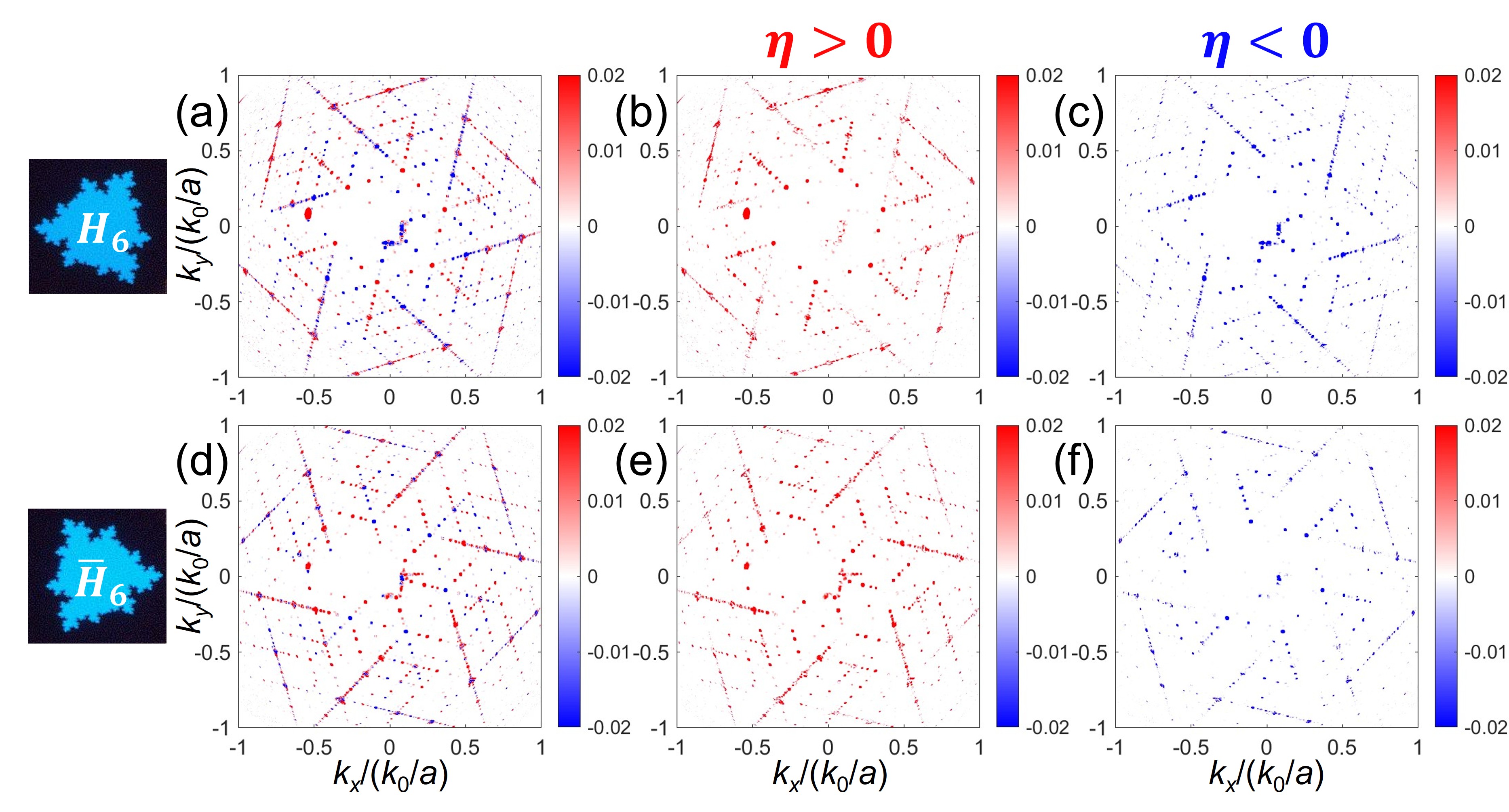}
\caption{\label{pol} $\eta$ maps, which show dependence of the handedness of the incidence circularly polarized light, in $k$ space for $a$ = 700 nm. (a)-(c) $H_{6}$ and (d)-(f) $\bar{H_{6}}$. (b) ((e)) and (c) ((e)) are positive and negative parts of (a) ((d)) the $\eta$, respectively.}
\end{figure*}

To further investigate the diffraction properties of the monotile lattice, we measured the diffraction pattern under circularly polarized incident light (See S4 in Supplementary Information for other polarization dependence).
Surprisingly, we discovered an intriguing dependence on the handedness of the incident circular polarization.
To evaluating the degree of circular polarization, we calculated
\begin{equation}
    \eta = \frac{I_{\text{LCP}}-I_{\text{RCP}}}{I_{\text{LCP}}+I_{\text{RCP}}}
\end{equation}
from measured diffraction patterns.
Here, $I_{\text{LCP}}$ and $I_{\text{RCP}}$ is the intensities of diffraction for left-handed and right-handed circular polarization, respectively.
Figures \ref{pol}(a) and (d) presents experimental results obtained by illuminating $H_{6}$ and $\bar{H}_{6}$ monotile lattices ($a = 700$ nm) with circularly polarized light, respectively.
In Figures \ref{pol}, a two-dimensional moving average has been applied using data from the surrounding 5×5 pixels to enhance the visibility of the red and blue peaks.
As shown in Figs.\ref{pol}(a) and (d), the diffraction pattern of the monotile lattice exhibits a intriguing dependence on circular polarization incidence.
In systems such as the honeycomb lattice and the Penrose tiling, which possess not only rotational symmetry but also inversion symmetry, Bragg peaks appear only at wave vectors that respect these symmetries. Consequently, due to this symmetry, no dependence on the handedness of the incident circular polarization emerges.
On the other hand, the monotile lattice possesses rotational symmetry but lacks inversion symmetry, allowing Bragg peaks to appear at various wave vectors. This absence of inversion symmetry enables these peaks to exhibit a dependence on the handedness of the incident circular polarization.
To make the complex distribution of the red and blue peaks more visible, the positive and negative regions of $\eta_{\text{chiral}}$ are plotted separately in Fig. \ref{pol}(b), (c), (e), and (f).
By comparing Figs. \ref{pol}(b) and (c), the shapes formed by the red and blue peaks are distinct from each other if we carefully observe the detail.
Moreover, these shapes are interchanged for the mirrored structure as shown in Fig. \ref{pol}(e) and (f).
The same features are also observed in other cases of $a$ (See S4 in Supplementary Information).
However, interestingly, the shapes of the red or blue peaks are different for each $a$.
This result indicates that the observed circular polarization dependence depends not only on the properties of the monotile itself but also on other parameters, such as the structure's periodicity and the diffraction angle.
This experimental result reveals a property unique to the monotile lattice that is not observed in conventional quasiperiodic systems with inversion symmetry.

In conclusion, we experimentally observed diffraction patterns from a monotile lattice generated from the hat tiling and reveal its unique chiral natures in $k$ space. 
The diffraction patterns exhibited sharp peaks that reflected the long-range order of the system, as well as symmetry features derived from the honeycomb frame.
We revealed the intriguing connection between the twisting of the metatiles and tiltiing of the diffraction patterns, inducing the chiral structure of the diffraction.
Moreover, the diffraction showed clear circular polarization dependence and exhibit interesting chiral properties.
This study represents the first experimental observation of the reciprocal nature of the hat tiling.
Our results reveal that the monotile lattice exhibits properties not found in conventional quasiperiodic structures, opening new frontier for the study of aperiodic structures.
%Unlike Penrose tiles and other quasicrystalline systems, the hat tile is a homogeneous system without a central point of symmetry.
%Our study showed photonic systems are useful for the experimental study of monotiles.
%An interesting open question for future research is whether the numerous satellite peaks observed in this study could contribute to the formation of a complete bandgap.

\begin{acknowledgments}
This work was supported by KAKENHI JP20H05641, JP21K14551, 24K01377, 24H02232 and
24H00400.
\end{acknowledgments}

\bibliography{b_monotile,b_photonic_quasicrystal,b_qp_meta,b_quasicrystal}% Produces the bibliography via BibTeX.

\providecommand{\noopsort}[1]{}\providecommand{\singleletter}[1]{#1}%
\begin{thebibliography}{29}%
\makeatletter
\providecommand \@ifxundefined [1]{%
 \@ifx{#1\undefined}
}%
\providecommand \@ifnum [1]{%
 \ifnum #1\expandafter \@firstoftwo
 \else \expandafter \@secondoftwo
 \fi
}%
\providecommand \@ifx [1]{%
 \ifx #1\expandafter \@firstoftwo
 \else \expandafter \@secondoftwo
 \fi
}%
\providecommand \natexlab [1]{#1}%
\providecommand \enquote  [1]{``#1''}%
\providecommand \bibnamefont  [1]{#1}%
\providecommand \bibfnamefont [1]{#1}%
\providecommand \citenamefont [1]{#1}%
\providecommand \href@noop [0]{\@secondoftwo}%
\providecommand \href [0]{\begingroup \@sanitize@url \@href}%
\providecommand \@href[1]{\@@startlink{#1}\@@href}%
\providecommand \@@href[1]{\endgroup#1\@@endlink}%
\providecommand \@sanitize@url [0]{\catcode `\\12\catcode `\$12\catcode `\&12\catcode `\#12\catcode `\^12\catcode `\_12\catcode `\%12\relax}%
\providecommand \@@startlink[1]{}%
\providecommand \@@endlink[0]{}%
\providecommand \url  [0]{\begingroup\@sanitize@url \@url }%
\providecommand \@url [1]{\endgroup\@href {#1}{\urlprefix }}%
\providecommand \urlprefix  [0]{URL }%
\providecommand \Eprint [0]{\href }%
\providecommand \doibase [0]{https://doi.org/}%
\providecommand \selectlanguage [0]{\@gobble}%
\providecommand \bibinfo  [0]{\@secondoftwo}%
\providecommand \bibfield  [0]{\@secondoftwo}%
\providecommand \translation [1]{[#1]}%
\providecommand \BibitemOpen [0]{}%
\providecommand \bibitemStop [0]{}%
\providecommand \bibitemNoStop [0]{.\EOS\space}%
\providecommand \EOS [0]{\spacefactor3000\relax}%
\providecommand \BibitemShut  [1]{\csname bibitem#1\endcsname}%
\let\auto@bib@innerbib\@empty
%</preamble>
\bibitem [{\citenamefont {Penrose}(1974)}]{Penrose1974}%
  \BibitemOpen
  \bibfield  {author} {\bibinfo {author} {\bibfnamefont {R.}~\bibnamefont {Penrose}},\ }\bibfield  {title} {\bibinfo {title} {The role of aesthetics in pure and applied mathematical research},\ }\href@noop {} {\bibfield  {journal} {\bibinfo  {journal} {Bulletin of the Institute of Mathematics and its Applications}\ }\textbf {\bibinfo {volume} {10}},\ \bibinfo {pages} {266} (\bibinfo {year} {1974})}\BibitemShut {NoStop}%
\bibitem [{\citenamefont {Suck}\ \emph {et~al.}(2002)\citenamefont {Suck}, \citenamefont {Schreiber},\ and\ \citenamefont {Häussler}}]{Suck2002}%
  \BibitemOpen
  \bibfield  {author} {\bibinfo {author} {\bibfnamefont {J.-B.}\ \bibnamefont {Suck}}, \bibinfo {author} {\bibfnamefont {M.}~\bibnamefont {Schreiber}},\ and\ \bibinfo {author} {\bibfnamefont {P.}~\bibnamefont {Häussler}},\ }\href@noop {} {\emph {\bibinfo {title} {Quasicrystals An Introduction to Structure, Physical Properties, ans Applications}}}\ (\bibinfo  {publisher} {Springer},\ \bibinfo {year} {2002})\BibitemShut {NoStop}%
\bibitem [{\citenamefont {Shechtman}\ \emph {et~al.}(1984)\citenamefont {Shechtman}, \citenamefont {Blech}, \citenamefont {Gratias},\ and\ \citenamefont {Cahn}}]{PhysRevLett.53.1951}%
  \BibitemOpen
  \bibfield  {author} {\bibinfo {author} {\bibfnamefont {D.}~\bibnamefont {Shechtman}}, \bibinfo {author} {\bibfnamefont {I.}~\bibnamefont {Blech}}, \bibinfo {author} {\bibfnamefont {D.}~\bibnamefont {Gratias}},\ and\ \bibinfo {author} {\bibfnamefont {J.~W.}\ \bibnamefont {Cahn}},\ }\bibfield  {title} {\bibinfo {title} {Metallic phase with long-range orientational order and no translational symmetry},\ }\href {https://doi.org/10.1103/PhysRevLett.53.1951} {\bibfield  {journal} {\bibinfo  {journal} {Phys. Rev. Lett.}\ }\textbf {\bibinfo {volume} {53}},\ \bibinfo {pages} {1951} (\bibinfo {year} {1984})}\BibitemShut {NoStop}%
\bibitem [{\citenamefont {Vardeny}\ \emph {et~al.}(2013)\citenamefont {Vardeny}, \citenamefont {Nahata},\ and\ \citenamefont {Agrawal}}]{Vardeny2013}%
  \BibitemOpen
  \bibfield  {author} {\bibinfo {author} {\bibfnamefont {Z.}~\bibnamefont {Vardeny}}, \bibinfo {author} {\bibfnamefont {A.}~\bibnamefont {Nahata}},\ and\ \bibinfo {author} {\bibfnamefont {A.}~\bibnamefont {Agrawal}},\ }\bibfield  {title} {\bibinfo {title} {Optics of photonic quasicrystals},\ }\href {https://doi.org/10.1038/nphoton.2012.343} {\bibfield  {journal} {\bibinfo  {journal} {Nature Photonics}\ }\textbf {\bibinfo {volume} {7}},\ \bibinfo {pages} {177} (\bibinfo {year} {2013})}\BibitemShut {NoStop}%
\bibitem [{\citenamefont {Steurer}\ and\ \citenamefont {Sutter-Widmer}(2007)}]{Steurer_2007}%
  \BibitemOpen
  \bibfield  {author} {\bibinfo {author} {\bibfnamefont {W.}~\bibnamefont {Steurer}}\ and\ \bibinfo {author} {\bibfnamefont {D.}~\bibnamefont {Sutter-Widmer}},\ }\bibfield  {title} {\bibinfo {title} {Photonic and phononic quasicrystals},\ }\href {https://doi.org/10.1088/0022-3727/40/13/R01} {\bibfield  {journal} {\bibinfo  {journal} {Journal of Physics D: Applied Physics}\ }\textbf {\bibinfo {volume} {40}},\ \bibinfo {pages} {R229} (\bibinfo {year} {2007})}\BibitemShut {NoStop}%
\bibitem [{\citenamefont {Maciá}(2012)}]{Maciá_2012}%
  \BibitemOpen
  \bibfield  {author} {\bibinfo {author} {\bibfnamefont {E.}~\bibnamefont {Maciá}},\ }\bibfield  {title} {\bibinfo {title} {Exploiting aperiodic designs in nanophotonic devices},\ }\href {https://doi.org/10.1088/0034-4885/75/3/036502} {\bibfield  {journal} {\bibinfo  {journal} {Reports on Progress in Physics}\ }\textbf {\bibinfo {volume} {75}},\ \bibinfo {pages} {036502} (\bibinfo {year} {2012})}\BibitemShut {NoStop}%
\bibitem [{\citenamefont {Jin}\ \emph {et~al.}(1999)\citenamefont {Jin}, \citenamefont {Cheng}, \citenamefont {Man}, \citenamefont {Li}, \citenamefont {Zhang}, \citenamefont {Ban},\ and\ \citenamefont {Sun}}]{10.1063/1.124848}%
  \BibitemOpen
  \bibfield  {author} {\bibinfo {author} {\bibfnamefont {C.}~\bibnamefont {Jin}}, \bibinfo {author} {\bibfnamefont {B.}~\bibnamefont {Cheng}}, \bibinfo {author} {\bibfnamefont {B.}~\bibnamefont {Man}}, \bibinfo {author} {\bibfnamefont {Z.}~\bibnamefont {Li}}, \bibinfo {author} {\bibfnamefont {D.}~\bibnamefont {Zhang}}, \bibinfo {author} {\bibfnamefont {S.}~\bibnamefont {Ban}},\ and\ \bibinfo {author} {\bibfnamefont {B.}~\bibnamefont {Sun}},\ }\bibfield  {title} {\bibinfo {title} {Band gap and wave guiding effect in a quasiperiodic photonic crystal},\ }\href {https://doi.org/10.1063/1.124848} {\bibfield  {journal} {\bibinfo  {journal} {Applied Physics Letters}\ }\textbf {\bibinfo {volume} {75}},\ \bibinfo {pages} {1848} (\bibinfo {year} {1999})}\BibitemShut {NoStop}%
\bibitem [{\citenamefont {Zoorob}\ \emph {et~al.}(2000)\citenamefont {Zoorob}, \citenamefont {Charlton}, \citenamefont {Parker} \emph {et~al.}}]{zoorob2000}%
  \BibitemOpen
  \bibfield  {author} {\bibinfo {author} {\bibfnamefont {M.}~\bibnamefont {Zoorob}}, \bibinfo {author} {\bibfnamefont {M.}~\bibnamefont {Charlton}}, \bibinfo {author} {\bibfnamefont {G.}~\bibnamefont {Parker}}, \emph {et~al.},\ }\bibfield  {title} {\bibinfo {title} {Complete photonic bandgaps in 12-fold symmetric quasicrystals},\ }\href {https://doi.org/10.1038/35008023} {\bibfield  {journal} {\bibinfo  {journal} {Nature}\ }\textbf {\bibinfo {volume} {404}},\ \bibinfo {pages} {740} (\bibinfo {year} {2000})}\BibitemShut {NoStop}%
\bibitem [{\citenamefont {Zhang}\ \emph {et~al.}(2001)\citenamefont {Zhang}, \citenamefont {Zhang},\ and\ \citenamefont {Chan}}]{PhysRevB.63.081105}%
  \BibitemOpen
  \bibfield  {author} {\bibinfo {author} {\bibfnamefont {X.}~\bibnamefont {Zhang}}, \bibinfo {author} {\bibfnamefont {Z.-Q.}\ \bibnamefont {Zhang}},\ and\ \bibinfo {author} {\bibfnamefont {C.~T.}\ \bibnamefont {Chan}},\ }\bibfield  {title} {\bibinfo {title} {Absolute photonic band gaps in 12-fold symmetric photonic quasicrystals},\ }\href {https://doi.org/10.1103/PhysRevB.63.081105} {\bibfield  {journal} {\bibinfo  {journal} {Phys. Rev. B}\ }\textbf {\bibinfo {volume} {63}},\ \bibinfo {pages} {081105} (\bibinfo {year} {2001})}\BibitemShut {NoStop}%
\bibitem [{\citenamefont {Rechtsman}\ \emph {et~al.}(2008)\citenamefont {Rechtsman}, \citenamefont {Jeong}, \citenamefont {Chaikin}, \citenamefont {Torquato},\ and\ \citenamefont {Steinhardt}}]{PhysRevLett.101.073902}%
  \BibitemOpen
  \bibfield  {author} {\bibinfo {author} {\bibfnamefont {M.~C.}\ \bibnamefont {Rechtsman}}, \bibinfo {author} {\bibfnamefont {H.-C.}\ \bibnamefont {Jeong}}, \bibinfo {author} {\bibfnamefont {P.~M.}\ \bibnamefont {Chaikin}}, \bibinfo {author} {\bibfnamefont {S.}~\bibnamefont {Torquato}},\ and\ \bibinfo {author} {\bibfnamefont {P.~J.}\ \bibnamefont {Steinhardt}},\ }\bibfield  {title} {\bibinfo {title} {Optimized structures for photonic quasicrystals},\ }\href {https://doi.org/10.1103/PhysRevLett.101.073902} {\bibfield  {journal} {\bibinfo  {journal} {Phys. Rev. Lett.}\ }\textbf {\bibinfo {volume} {101}},\ \bibinfo {pages} {073902} (\bibinfo {year} {2008})}\BibitemShut {NoStop}%
\bibitem [{\citenamefont {Notomi}\ \emph {et~al.}(2004)\citenamefont {Notomi}, \citenamefont {Suzuki}, \citenamefont {Tamamura},\ and\ \citenamefont {Edagawa}}]{PhysRevLett.92.123906}%
  \BibitemOpen
  \bibfield  {author} {\bibinfo {author} {\bibfnamefont {M.}~\bibnamefont {Notomi}}, \bibinfo {author} {\bibfnamefont {H.}~\bibnamefont {Suzuki}}, \bibinfo {author} {\bibfnamefont {T.}~\bibnamefont {Tamamura}},\ and\ \bibinfo {author} {\bibfnamefont {K.}~\bibnamefont {Edagawa}},\ }\bibfield  {title} {\bibinfo {title} {Lasing action due to the two-dimensional quasiperiodicity of photonic quasicrystals with a penrose lattice},\ }\href {https://doi.org/10.1103/PhysRevLett.92.123906} {\bibfield  {journal} {\bibinfo  {journal} {Phys. Rev. Lett.}\ }\textbf {\bibinfo {volume} {92}},\ \bibinfo {pages} {123906} (\bibinfo {year} {2004})}\BibitemShut {NoStop}%
\bibitem [{\citenamefont {Nozaki}\ and\ \citenamefont {Baba}(2004)}]{10.1063/1.1762705}%
  \BibitemOpen
  \bibfield  {author} {\bibinfo {author} {\bibfnamefont {K.}~\bibnamefont {Nozaki}}\ and\ \bibinfo {author} {\bibfnamefont {T.}~\bibnamefont {Baba}},\ }\bibfield  {title} {\bibinfo {title} {Quasiperiodic photonic crystal microcavity lasers},\ }\href@noop {} {\bibfield  {journal} {\bibinfo  {journal} {Applied Physics Letters}\ }\textbf {\bibinfo {volume} {84}},\ \bibinfo {pages} {4875} (\bibinfo {year} {2004})}\BibitemShut {NoStop}%
\bibitem [{\citenamefont {Mahler}\ \emph {et~al.}(2010)\citenamefont {Mahler}, \citenamefont {Tredicucci}, \citenamefont {Beltram}, \citenamefont {Walther}, \citenamefont {Faist}, \citenamefont {Beere}, \citenamefont {Ritchie},\ and\ \citenamefont {Wiersma}}]{Mahler2010}%
  \BibitemOpen
  \bibfield  {author} {\bibinfo {author} {\bibfnamefont {L.}~\bibnamefont {Mahler}}, \bibinfo {author} {\bibfnamefont {A.}~\bibnamefont {Tredicucci}}, \bibinfo {author} {\bibfnamefont {F.}~\bibnamefont {Beltram}}, \bibinfo {author} {\bibfnamefont {C.}~\bibnamefont {Walther}}, \bibinfo {author} {\bibfnamefont {J.}~\bibnamefont {Faist}}, \bibinfo {author} {\bibfnamefont {H.~E.}\ \bibnamefont {Beere}}, \bibinfo {author} {\bibfnamefont {D.~A.}\ \bibnamefont {Ritchie}},\ and\ \bibinfo {author} {\bibfnamefont {D.~S.}\ \bibnamefont {Wiersma}},\ }\bibfield  {title} {\bibinfo {title} {Quasi-periodic distributed feedback laser},\ }\href {https://doi.org/10.1038/nphoton.2009.285} {\bibfield  {journal} {\bibinfo  {journal} {Nature Photonics}\ }\textbf {\bibinfo {volume} {4}},\ \bibinfo {pages} {165} (\bibinfo {year} {2010})}\BibitemShut {NoStop}%
\bibitem [{\citenamefont {Vitiello}\ \emph {et~al.}(2014)\citenamefont {Vitiello}, \citenamefont {Nobile}, \citenamefont {Ronzani}, \citenamefont {Tredicucci}, \citenamefont {Castellano}, \citenamefont {Talora}, \citenamefont {Li}, \citenamefont {Linfield},\ and\ \citenamefont {Davies}}]{Vitiello2014}%
  \BibitemOpen
  \bibfield  {author} {\bibinfo {author} {\bibfnamefont {M.~S.}\ \bibnamefont {Vitiello}}, \bibinfo {author} {\bibfnamefont {M.}~\bibnamefont {Nobile}}, \bibinfo {author} {\bibfnamefont {A.}~\bibnamefont {Ronzani}}, \bibinfo {author} {\bibfnamefont {A.}~\bibnamefont {Tredicucci}}, \bibinfo {author} {\bibfnamefont {F.}~\bibnamefont {Castellano}}, \bibinfo {author} {\bibfnamefont {V.}~\bibnamefont {Talora}}, \bibinfo {author} {\bibfnamefont {L.}~\bibnamefont {Li}}, \bibinfo {author} {\bibfnamefont {E.~H.}\ \bibnamefont {Linfield}},\ and\ \bibinfo {author} {\bibfnamefont {A.~G.}\ \bibnamefont {Davies}},\ }\bibfield  {title} {\bibinfo {title} {Photonic quasi-crystal terahertz lasers},\ }\href {https://doi.org/10.1038/ncomms6884} {\bibfield  {journal} {\bibinfo  {journal} {Nature Communications}\ }\textbf {\bibinfo {volume} {5}},\ \bibinfo {pages} {5884} (\bibinfo {year} {2014})}\BibitemShut {NoStop}%
\bibitem [{\citenamefont {Yulevich}\ \emph {et~al.}(2015)\citenamefont {Yulevich}, \citenamefont {Maguid}, \citenamefont {Shitrit}, \citenamefont {Veksler}, \citenamefont {Kleiner},\ and\ \citenamefont {Hasman}}]{PhysRevLett.115.205501}%
  \BibitemOpen
  \bibfield  {author} {\bibinfo {author} {\bibfnamefont {I.}~\bibnamefont {Yulevich}}, \bibinfo {author} {\bibfnamefont {E.}~\bibnamefont {Maguid}}, \bibinfo {author} {\bibfnamefont {N.}~\bibnamefont {Shitrit}}, \bibinfo {author} {\bibfnamefont {D.}~\bibnamefont {Veksler}}, \bibinfo {author} {\bibfnamefont {V.}~\bibnamefont {Kleiner}},\ and\ \bibinfo {author} {\bibfnamefont {E.}~\bibnamefont {Hasman}},\ }\bibfield  {title} {\bibinfo {title} {Optical mode control by geometric phase in quasicrystal metasurface},\ }\href {https://doi.org/10.1103/PhysRevLett.115.205501} {\bibfield  {journal} {\bibinfo  {journal} {Phys. Rev. Lett.}\ }\textbf {\bibinfo {volume} {115}},\ \bibinfo {pages} {205501} (\bibinfo {year} {2015})}\BibitemShut {NoStop}%
\bibitem [{\citenamefont {Xu}\ \emph {et~al.}(2024)\citenamefont {Xu}, \citenamefont {Zhao}, \citenamefont {Zhang} \emph {et~al.}}]{xu2024}%
  \BibitemOpen
  \bibfield  {author} {\bibinfo {author} {\bibfnamefont {C.}~\bibnamefont {Xu}}, \bibinfo {author} {\bibfnamefont {R.}~\bibnamefont {Zhao}}, \bibinfo {author} {\bibfnamefont {X.}~\bibnamefont {Zhang}}, \emph {et~al.},\ }\bibfield  {title} {\bibinfo {title} {Quasicrystal metasurface for dual functionality of holography and diffraction generation},\ }\href {https://doi.org/10.1186/s43593-024-00065-7} {\bibfield  {journal} {\bibinfo  {journal} {eLight}\ }\textbf {\bibinfo {volume} {4}},\ \bibinfo {pages} {9} (\bibinfo {year} {2024})}\BibitemShut {NoStop}%
\bibitem [{\citenamefont {Arjas}\ \emph {et~al.}(2024)\citenamefont {Arjas}, \citenamefont {Taskinen}, \citenamefont {Heilmann} \emph {et~al.}}]{arjas2024}%
  \BibitemOpen
  \bibfield  {author} {\bibinfo {author} {\bibfnamefont {K.}~\bibnamefont {Arjas}}, \bibinfo {author} {\bibfnamefont {J.~M.}\ \bibnamefont {Taskinen}}, \bibinfo {author} {\bibfnamefont {R.}~\bibnamefont {Heilmann}}, \emph {et~al.},\ }\bibfield  {title} {\bibinfo {title} {High topological charge lasing in quasicrystals},\ }\href {https://doi.org/10.1038/s41467-024-53952-5} {\bibfield  {journal} {\bibinfo  {journal} {Nature Communications}\ }\textbf {\bibinfo {volume} {15}},\ \bibinfo {pages} {9544} (\bibinfo {year} {2024})}\BibitemShut {NoStop}%
\bibitem [{\citenamefont {Kraus}\ \emph {et~al.}(2012)\citenamefont {Kraus}, \citenamefont {Lahini}, \citenamefont {Ringel}, \citenamefont {Verbin},\ and\ \citenamefont {Zilberberg}}]{PhysRevLett.109.106402}%
  \BibitemOpen
  \bibfield  {author} {\bibinfo {author} {\bibfnamefont {Y.~E.}\ \bibnamefont {Kraus}}, \bibinfo {author} {\bibfnamefont {Y.}~\bibnamefont {Lahini}}, \bibinfo {author} {\bibfnamefont {Z.}~\bibnamefont {Ringel}}, \bibinfo {author} {\bibfnamefont {M.}~\bibnamefont {Verbin}},\ and\ \bibinfo {author} {\bibfnamefont {O.}~\bibnamefont {Zilberberg}},\ }\bibfield  {title} {\bibinfo {title} {Topological states and adiabatic pumping in quasicrystals},\ }\href {https://doi.org/10.1103/PhysRevLett.109.106402} {\bibfield  {journal} {\bibinfo  {journal} {Phys. Rev. Lett.}\ }\textbf {\bibinfo {volume} {109}},\ \bibinfo {pages} {106402} (\bibinfo {year} {2012})}\BibitemShut {NoStop}%
\bibitem [{\citenamefont {Smith}\ \emph {et~al.}(2024{\natexlab{a}})\citenamefont {Smith}, \citenamefont {Myers}, \citenamefont {Kaplan},\ and\ \citenamefont {Goodman-Strauss}}]{smith2024}%
  \BibitemOpen
  \bibfield  {author} {\bibinfo {author} {\bibfnamefont {D.}~\bibnamefont {Smith}}, \bibinfo {author} {\bibfnamefont {J.~S.}\ \bibnamefont {Myers}}, \bibinfo {author} {\bibfnamefont {C.~S.}\ \bibnamefont {Kaplan}},\ and\ \bibinfo {author} {\bibfnamefont {C.}~\bibnamefont {Goodman-Strauss}},\ }\bibfield  {title} {\bibinfo {title} {An aperiodic monotile},\ }\bibfield  {journal} {\bibinfo  {journal} {Combinatorial Theory}\ }\textbf {\bibinfo {volume} {4}},\ \href {https://doi.org/10.5070/C64163843} {10.5070/C64163843} (\bibinfo {year} {2024}{\natexlab{a}})\BibitemShut {NoStop}%
\bibitem [{\citenamefont {Schirmann}\ \emph {et~al.}(2024)\citenamefont {Schirmann}, \citenamefont {Franca}, \citenamefont {Flicker},\ and\ \citenamefont {Grushin}}]{PhysRevLett.132.086402}%
  \BibitemOpen
  \bibfield  {author} {\bibinfo {author} {\bibfnamefont {J.}~\bibnamefont {Schirmann}}, \bibinfo {author} {\bibfnamefont {S.}~\bibnamefont {Franca}}, \bibinfo {author} {\bibfnamefont {F.}~\bibnamefont {Flicker}},\ and\ \bibinfo {author} {\bibfnamefont {A.~G.}\ \bibnamefont {Grushin}},\ }\bibfield  {title} {\bibinfo {title} {Physical properties of an aperiodic monotile with graphene-like features, chirality, and zero modes},\ }\href {https://doi.org/10.1103/PhysRevLett.132.086402} {\bibfield  {journal} {\bibinfo  {journal} {Phys. Rev. Lett.}\ }\textbf {\bibinfo {volume} {132}},\ \bibinfo {pages} {086402} (\bibinfo {year} {2024})}\BibitemShut {NoStop}%
\bibitem [{\citenamefont {Kaplan}\ \emph {et~al.}(2024)\citenamefont {Kaplan}, \citenamefont {O'Keeffe},\ and\ \citenamefont {Treacy}}]{Kaplan:uv5021}%
  \BibitemOpen
  \bibfield  {author} {\bibinfo {author} {\bibfnamefont {C.~S.}\ \bibnamefont {Kaplan}}, \bibinfo {author} {\bibfnamefont {M.}~\bibnamefont {O'Keeffe}},\ and\ \bibinfo {author} {\bibfnamefont {M.~M.~J.}\ \bibnamefont {Treacy}},\ }\bibfield  {title} {\bibinfo {title} {{Periodic diffraction from an aperiodic monohedral tiling}},\ }\href {https://doi.org/10.1107/S2053273323009506} {\bibfield  {journal} {\bibinfo  {journal} {Acta Crystallographica Section A}\ }\textbf {\bibinfo {volume} {80}},\ \bibinfo {pages} {72} (\bibinfo {year} {2024})}\BibitemShut {NoStop}%
\bibitem [{\citenamefont {Okabe}\ \emph {et~al.}(2024)\citenamefont {Okabe}, \citenamefont {Niizeki},\ and\ \citenamefont {Araki}}]{Okabe_2024}%
  \BibitemOpen
  \bibfield  {author} {\bibinfo {author} {\bibfnamefont {Y.}~\bibnamefont {Okabe}}, \bibinfo {author} {\bibfnamefont {K.}~\bibnamefont {Niizeki}},\ and\ \bibinfo {author} {\bibfnamefont {Y.}~\bibnamefont {Araki}},\ }\bibfield  {title} {\bibinfo {title} {Ising model on the aperiodic smith hat},\ }\href {https://doi.org/10.1088/1751-8121/ad2f70} {\bibfield  {journal} {\bibinfo  {journal} {Journal of Physics A: Mathematical and Theoretical}\ }\textbf {\bibinfo {volume} {57}},\ \bibinfo {pages} {125004} (\bibinfo {year} {2024})}\BibitemShut {NoStop}%
\bibitem [{\citenamefont {Jin}\ \emph {et~al.}(2024)\citenamefont {Jin}, \citenamefont {Ni}, \citenamefont {Chu}, \citenamefont {Sun}, \citenamefont {Xu}, \citenamefont {Hou}, \citenamefont {Hou}, \citenamefont {Zhong}, \citenamefont {Liu},\ and\ \citenamefont {Xiong}}]{JIN2024130781}%
  \BibitemOpen
  \bibfield  {author} {\bibinfo {author} {\bibfnamefont {X.}~\bibnamefont {Jin}}, \bibinfo {author} {\bibfnamefont {B.}~\bibnamefont {Ni}}, \bibinfo {author} {\bibfnamefont {G.}~\bibnamefont {Chu}}, \bibinfo {author} {\bibfnamefont {C.}~\bibnamefont {Sun}}, \bibinfo {author} {\bibfnamefont {B.}~\bibnamefont {Xu}}, \bibinfo {author} {\bibfnamefont {L.}~\bibnamefont {Hou}}, \bibinfo {author} {\bibfnamefont {J.~J.}\ \bibnamefont {Hou}}, \bibinfo {author} {\bibfnamefont {C.}~\bibnamefont {Zhong}}, \bibinfo {author} {\bibfnamefont {X.}~\bibnamefont {Liu}},\ and\ \bibinfo {author} {\bibfnamefont {J.}~\bibnamefont {Xiong}},\ }\bibfield  {title} {\bibinfo {title} {Silicon artificial neurons for uniform signal transmission and amplification},\ }\href {https://doi.org/https://doi.org/10.1016/j.optcom.2024.130781} {\bibfield  {journal} {\bibinfo  {journal} {Optics Communications}\ }\textbf {\bibinfo {volume} {569}},\ \bibinfo {pages} {130781} (\bibinfo {year} {2024})}\BibitemShut {NoStop}%
\bibitem [{\citenamefont {Xiao}\ and\ \citenamefont {Chan}(2024)}]{PhysRevB.110.075435}%
  \BibitemOpen
  \bibfield  {author} {\bibinfo {author} {\bibfnamefont {Y.-X.}\ \bibnamefont {Xiao}}\ and\ \bibinfo {author} {\bibfnamefont {C.~T.}\ \bibnamefont {Chan}},\ }\bibfield  {title} {\bibinfo {title} {Obstacle-insensitive eigenfields due to boundary condition--symmetry compatibility},\ }\href {https://doi.org/10.1103/PhysRevB.110.075435} {\bibfield  {journal} {\bibinfo  {journal} {Phys. Rev. B}\ }\textbf {\bibinfo {volume} {110}},\ \bibinfo {pages} {075435} (\bibinfo {year} {2024})}\BibitemShut {NoStop}%
\bibitem [{\citenamefont {Singh}\ and\ \citenamefont {Flicker}(2024)}]{PhysRevB.109.L220303}%
  \BibitemOpen
  \bibfield  {author} {\bibinfo {author} {\bibfnamefont {S.}~\bibnamefont {Singh}}\ and\ \bibinfo {author} {\bibfnamefont {F.}~\bibnamefont {Flicker}},\ }\bibfield  {title} {\bibinfo {title} {Exact solution to the quantum and classical dimer models on the spectre aperiodic monotiling},\ }\href {https://doi.org/10.1103/PhysRevB.109.L220303} {\bibfield  {journal} {\bibinfo  {journal} {Phys. Rev. B}\ }\textbf {\bibinfo {volume} {109}},\ \bibinfo {pages} {L220303} (\bibinfo {year} {2024})}\BibitemShut {NoStop}%
\bibitem [{\citenamefont {Sun}\ and\ \citenamefont {Guo}(2024)}]{202407.1532}%
  \BibitemOpen
  \bibfield  {author} {\bibinfo {author} {\bibfnamefont {B.~H.}\ \bibnamefont {Sun}}\ and\ \bibinfo {author} {\bibfnamefont {X.~L.}\ \bibnamefont {Guo}},\ }\bibfield  {title} {\bibinfo {title} {Mechanics of topological high-entropy structures made out of ‘einstein’ puzzle pieces},\ }\bibfield  {journal} {\bibinfo  {journal} {Preprints}\ }\href {https://doi.org/10.20944/preprints202407.1532.v1} {10.20944/preprints202407.1532.v1} (\bibinfo {year} {2024})\BibitemShut {NoStop}%
\bibitem [{\citenamefont {Clarke}\ \emph {et~al.}(2023)\citenamefont {Clarke}, \citenamefont {Carter}, \citenamefont {Jowers},\ and\ \citenamefont {Moat}}]{CLARKE2023101959}%
  \BibitemOpen
  \bibfield  {author} {\bibinfo {author} {\bibfnamefont {D.~J.}\ \bibnamefont {Clarke}}, \bibinfo {author} {\bibfnamefont {F.}~\bibnamefont {Carter}}, \bibinfo {author} {\bibfnamefont {I.}~\bibnamefont {Jowers}},\ and\ \bibinfo {author} {\bibfnamefont {R.~J.}\ \bibnamefont {Moat}},\ }\bibfield  {title} {\bibinfo {title} {An isotropic zero poisson's ratio metamaterial based on the aperiodic ‘hat’ monotile},\ }\href {https://doi.org/https://doi.org/10.1016/j.apmt.2023.101959} {\bibfield  {journal} {\bibinfo  {journal} {Applied Materials Today}\ }\textbf {\bibinfo {volume} {35}},\ \bibinfo {pages} {101959} (\bibinfo {year} {2023})}\BibitemShut {NoStop}%
\bibitem [{\citenamefont {Smith}\ \emph {et~al.}(2024{\natexlab{b}})\citenamefont {Smith}, \citenamefont {Myers}, \citenamefont {Kaplan},\ and\ \citenamefont {Goodman-Strauss}}]{Smith2024b}%
  \BibitemOpen
  \bibfield  {author} {\bibinfo {author} {\bibfnamefont {D.}~\bibnamefont {Smith}}, \bibinfo {author} {\bibfnamefont {J.~S.}\ \bibnamefont {Myers}}, \bibinfo {author} {\bibfnamefont {C.~S.}\ \bibnamefont {Kaplan}},\ and\ \bibinfo {author} {\bibfnamefont {C.}~\bibnamefont {Goodman-Strauss}},\ }\bibfield  {title} {\bibinfo {title} {A chiral aperiodic monotile},\ }\bibfield  {journal} {\bibinfo  {journal} {Combinatorial Theory}\ }\textbf {\bibinfo {volume} {4}},\ \href {https://doi.org/10.5070/C64264241} {10.5070/C64264241} (\bibinfo {year} {2024}{\natexlab{b}})\BibitemShut {NoStop}%
\bibitem [{\citenamefont {Pan}\ \emph {et~al.}(2023)\citenamefont {Pan}, \citenamefont {Sun}, \citenamefont {Zhang}, \citenamefont {Li}, \citenamefont {Kong}, \citenamefont {Zhao},\ and\ \citenamefont {Gao}}]{photonics10091045}%
  \BibitemOpen
  \bibfield  {author} {\bibinfo {author} {\bibfnamefont {Y.}~\bibnamefont {Pan}}, \bibinfo {author} {\bibfnamefont {X.-F.}\ \bibnamefont {Sun}}, \bibinfo {author} {\bibfnamefont {G.-B.}\ \bibnamefont {Zhang}}, \bibinfo {author} {\bibfnamefont {Q.-L.}\ \bibnamefont {Li}}, \bibinfo {author} {\bibfnamefont {Y.-N.}\ \bibnamefont {Kong}}, \bibinfo {author} {\bibfnamefont {T.-F.}\ \bibnamefont {Zhao}},\ and\ \bibinfo {author} {\bibfnamefont {X.-Z.}\ \bibnamefont {Gao}},\ }\bibfield  {title} {\bibinfo {title} {Two-dimensional quasi-periodic diffraction properties of the scalar and vector optical fields},\ }\bibfield  {journal} {\bibinfo  {journal} {Photonics}\ }\textbf {\bibinfo {volume} {10}},\ \href {https://doi.org/10.3390/photonics10091045} {10.3390/photonics10091045} (\bibinfo {year} {2023})\BibitemShut {NoStop}%
\end{thebibliography}%

\end{document}